\documentclass[preprint,onecolumn,
 amsmath,amssymb,
 aps,
]{revtex4-1}
\usepackage{graphicx,color}
\usepackage{dcolumn}
\usepackage{bm}

\begin{document}

\title{On the Possible Evolutionary History of the Water Ocean on Venus}
\author{Tetsuya Hara}
\email{hara@cc.kyoto-su.ac.jp}
\affiliation{Department of Physics, Kyoto Sangyo University, Kyoto 603-8555, Japan}
\author{Anna Suzuki}
\email{i1785064@cc.kyoto-su.ac.jp}
\affiliation{Department of Physics, Kyoto Sangyo University, Kyoto 603-8555, Japan}

\begin{abstract}
We have investigated the possible evolutional history of the water ocean on Venus, adopting the one-dimensional radiative-convective model,
 including the parameters as albedo and relative humidity.
Under this model, it has the possibility that the habitable zone could include Venus.  
It could continue for $\sim 1$ Gy in faint young solar flux increasing,
 with modest parameters such as albedo = 0.3, relative humidity (RH=1), and $p_{n0}=10^5 $Pa.  If we relax parameters 
considering the 3-Dimensional calculations, the ocean could exist there longer than  $\sim$ 4.6 Gy.  
 In such cases, we have to consider the cause of runaway other than just solar luminosity increasing. 
 It is important to investigate Venus history for the coming future of Earth and 
 observations of exoplanets for their historical habitable zones.
 \end{abstract}
\maketitle


\section{Introduction}
One of the main problems of exoplanet science is to find out the habitable zone (Yang et al. 2014; Kane et al. 2019; Way \& Del Genio, 2020), where liquid water could exist.  
Venus is similar to Earth in size and position in our solar system (Taylor \& Grinspoon, 2009). 
Although it is not certain whether Venus ever had an ocean, it is investigated in this paper that a habitable period could be possible under the traditional estimate. 

It is discussed that water was abundant on Venus from the observation of excess deuterium concentration compared to that on Earth (Donahue et al. 1982).
 It is indicated from the observations by the Venus Express that the planet's highland plateaus were ancient continents surrounded by water 
and made of substances similar to granite (Hashimoto \& Sugiya, 2003; Hashimoto et al. 2008).  It must have happened a runaway greenhouse on Venus. 
 We have estimated the atmospheric structure on a planet in a simplified model, including the parameters as albedo and relative humidity.
 Under one dimensional radiative-convective model (Nakajima, Hayashi \& Abe, 1992), the inner habitable zones in the faint young solar flux could include Venus.  

 Hamano Abe \& Genda (2013) investigated the initial evolution of the atmosphere and magma ocean. They derived the emergence of two types of the terrestrial
planets on the solidification of the magma ocean.  Type I is the planets where magma solidifies within several $10^6$ years.  
Their distance from the central star is greater than 0.7 AU and the mass content of water is comparable and/or greater than that of Earth.
On the other hand, Type II is the planet where magma solidifies with more than $10^8$ years.  Their distance from the central star is smaller than 0.7 AU 
and the mass content of water is much smaller than that of Earth.  They pointed out the possibility that Venus may belong to Type II planets, however,  they did not determine.

If Venus was formed in a stage before the gas of a solar nebula was dissipated it was surrounded by a dense primitive atmosphere 
with a solar chemical composition.  Then the solar nebulae were dissipated  by strong solar wind in a T Tauri stage 
in an order of $10^6 \sim 10^7$ y (Sekiya, Nakazawa \& Hayashi 1981).  
It could be assumed that the magma ocean cools down relatively slow in an order of $\sim 10^8$ y (Harrison, 2009; Hamano et al. 2013)), 
most of the water vapor in the atmosphere fell as heavy rain and filled basins in the ground on Venus. 
After that, the Late Heavy Bombardment (LHB) might have occurred approximately 4.1 to 3.8 billion years (Gy) ago, as the same for Moon and Earth
(Hartmann, Quantinb \& Mangold 2007).  Although LHB scenario is questioned recently (Way \& Del Genio. 2020), we investigate the ocean history on Venus further.    
In Sect. II, the method of the model is outlined, and it is applied to Venus in Sect. III.  In Sect. IV, results, and discussions are deployed.

\section{One-Dimensional Radiative-Convective Model}
We have followed the method of Nakajima et al. (1992) for the One-Dimensional Radiative-Convective Model.  
The atmosphere is assumed to consist of the non-condensable component (mainly N$_2$) and condensable component (H$_2$O).  
 For the stratosphere, radiative equilibrium is assumed. 
 About the radiative transfer, the atmosphere is considered to be transparent to solar radiation in the optical range.  In the infrared range, 
the absorption coefficient is taken to be constant and independent of wavelength which is said to be the gray atmosphere. 
The radiation transfer equation is integrated by using the Eddington approximation.

Under the tropopause, the adiabatic lapse rate is adopted where water vapor is saturated. 
  It is assumed that the saturation water vapor pressure  $p^{*}$ is derived under the Clausius-Clapeyron relationship and given by
  \begin{equation}
     p^*(T)=p^{*}_o \exp \left( -\frac{l}{RT} \right),
  \end{equation}
where $ T$ and $R$ are temperature, the gas constant, respectively. 
The $l$, and $p^*_o$ are the latent heat of condensable component ($l=43655($J mol$^{-1}$) ), and the constant for water saturation curve ($p^{*}_o=1.4 \times 10^{11} $(Pa)).

    For the troposphere, it is assumed to be pseudoadiabatic lapse rate for the temperature gradient with pressure as
\begin{equation}
    \left(\frac{\partial T}{\partial p} \right)=\frac{\frac{RT}{pc_{pn}}+\frac{x_v^*}{x_n}\frac{l}{pc_{pn}}}{x_n+x_v^*\frac{c_{pn}}{c_{pn}}+\frac{x_v}{x_n}\frac{l^2}{RT^2c_{pn}}}
\end{equation}
where  $c_{pv}$, and $c_{pn}$, are the mole specific heat at constant pressure of condensable and noncondensable components, respectively.
The parameter $x_n$ and $x_v$ are the mole fractions of the saturation condensable and noncondensable components, respectively. 

   Given the noncondensable pressure $p_{n0}$ and temperature at the surface, $T(p)$ is obtained up to the atmosphere.  There appears to be a height where the 
net convergence becomes positive in the upper levels of the atmosphere. The position of the tropopause is taken to be there.

\section{Applied to Venus}
The first atmosphere would have consisted of the solar nebula, primarily H and He, which would be driven off by the solar wind.  
It is almost the same for the inner planets like Mercury, Earth, and Mars.
Outgassing from volcanism produced the next atmosphere, consisting largely of N$_2$, H$_2$O, and CO$_2$. 
If the atmosphere cools down enough, it rains and water ocean is formed.  Then CO$_2$ is soon dissolved in water and built up carbonate sediments.  
To some extent, it is the same with Earth and maybe Mars.
The time scale of the dissolution is of an order of $10^6 \sim 10^7$y.  At present, the rate of CO$_2$ dissolution into the sea
is 3.3 $\times 10^{17}$g per year (IPCC 2015) on Earth. 
If we assume almost the same environment for Venus, the dissolution time is $4.6 \times 10^{23}/( 3.3 \times 10^{17}) \sim 10^6$y, 
where we take the CO$_2$ quantity on Venus is $\sim 10^{23}$g on Venus (Way et al. 2016).   
Although the observed N$_2$ component is greater than that on Earth, it is taken that the non-condensable component (N$_2$) is about 1 atm ($ \sim 10^5$ Pa) 
for the moment.

We have applied the method to Venus, where the main differences are the solar constant (1364 W/m$^2$ for Earth and 2622W/m$^2$ for Venus) and
 the gravitational acceleration from 9.8m/s$^2$ to 8.87m/s$^2$. 
 Taking the spherical mean for the injected flux ($\times$ 1/4), solar flux is 341W/m$^2$ and 656W/m$^2$ for Earth and Venus, respectively.  
 As the solar luminosity is almost 70$\%$ of present value at 4.6Gy ago (Bahcall, Pinsonneault, \& Basu 2001), the flux is 239W/m$^2$ and 459W/m$^2$ for Earth and Venus, respectively.
 If we take the albedo $\sim$ 0.3 which is the corresponding value of Earth at present, the solar injection flux decreases to 167W/m$^2$ and 321W/m$^2$, respectively. 
 From the numerical results which are presented in Fig. 1, the critical solar flux for the greenhouse runaway on Venus is $\sim$345W/m$^2$,

Adopting these results, we have inferred the history of the water ocean on Venus. 
The solar luminosity increase each 10$^8$ y
 by 0.778 $\%$, 
because it increase 1.428 $\sim$ 1/0.7 times from the initial time (4.6 Gy ago) by calculating $(1.00778)^{46} \sim$ 1.428.
Then it takes almost 1 Gy that the solar flux increases from 321W/m$^2$ to 345W/m$^2$, 
using the calculation $(1.00778)^{9.3} \sim 345/321 \simeq 1.0748.$ 
Under these crude estimations, the ocean on Venus could exist continuously $\sim 1 (\sim 0.93)$ Gy until the solar flux increased to the critical solar flux of the greenhouse runaway.

From the numerical simulations of the 3-D calculation (Leconte et al. 2013) , 
the effective albedo shows the tendency of a significant increase.  
It is also insisted by the numerical simulation (Yang et al. 2014),  
that the planetary rotation rate strongly affects the inner edge of the habitable zone. 
Venus seems to be almost locked to Sun.
Considering those simulations, if we take the albedo of Venus as 0.5, 
the solar injection flux will become 230W/m$^2$.  
Then it will take $\sim$ 5.2 Gy to become the greenhouse runaway value, because of (1.00778)$^{52.3} \sim 345/230=1.5$.  
For a case longer than $\sim$ 4.6 Gy, there must be another cause for greenhouse runaway such as huge volcanic emission 
with global lava resurfacing event, pointed by Way \& Del Genio (2020).

In Fig. 1, the relationship between the surface temperature $T_s$ and solar flux (W/m$^2$) is presented. 
For the equilibrium, solar flux must be equal to the upward infrared radiation flux density from the top of the atmosphere $F_{IRtop}$.
The radiation fluxes 321W/m$^2$ for albedo = 0.3 and 230W/m$^2$ for albedo = 0.5 are indicated.  
The time between the initial planet formation and the runaway is estimated for each case.   The runaway flux $\sim$345W/m$^2$ is also shown.

   The stratosphere is in radiative equilibrium.  It is assumed that the mole fraction of the condensable component in the stratosphere is constant and is equal to the value at the tropopause (Nakajima et al. 1992).  This assumption increases the water vapor and the optical depth of the stratosphere which has caused the decrease of the outgoing radiation flux.  
   The green curve in Fig. 1  shows the case, not considering the decrease of the radiation flux due to the opacity increase of the stratosphere 
   (Nakajima et al. 1992; Seager 2010)

     
\begin{figure}[ht]
\includegraphics[clip,width=10cm,height=7cm]{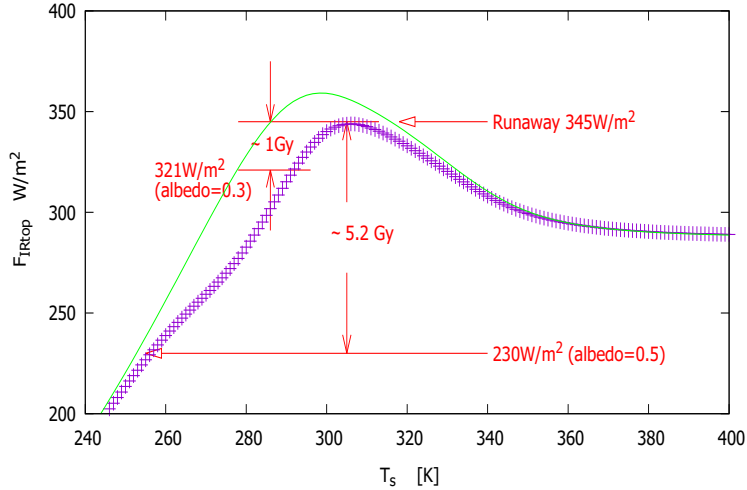}
\caption{The relationship between $T_s$ and $F_{IRtop}$. 
The solar radiation fluxes 321W/m$^2$ for albedo 0.3 and 230W/m$^2$ for albedo 0.5 are indicated.  
The time between the initial time (4.6Gy ago) and the runaway is estimated for each case.   The runaway flux density $\sim$345W/m$^2$ is also shown.
The green curve shows the case, not considering the decrease of the radiation flux due to the opacity increase of the stratosphere.
}   
\end{figure}
     


 
 At present, the N$_2$ mass in Venus atmosphere ($1.1\times 10^{19}$kg) is almost three times that of Earth  ($3.2\times 10^{18}$kg) (Way et al. 2016).
Then we have simulated three pressure cases of non-condensable component (N$_2$) ($p_{n0}=10^5, 3\times 10^5, $ and $10^6$ Pa) which are presented in Fig. 2. 
Increasing the value of $p_{n0}$, the runaway critical solar flux increases from 345W/m$^2$ ($p_{n0}=10^5$ Pa) to 360 W/m$^2$ ($p_{n0}=3\times 10^5$ Pa) 
and 374W/m$^2$ ($p_{n0}=10^6 $ Pa), shown there.  The ocean could exist $\sim$1Gy , $\sim $1.5 Gy and $\sim$2.0 Gy, 
because of $(1.00778)^{14.8} \sim 360/321 \simeq 1.1215$, and $(1.00778)^{19.7} \sim 374/321 \simeq 1.1651$, respectively.  
If we consider cases of albedo = 0.5, the ocean could exist longer than $\sim$ 4.6 Gy.

\begin{figure}[ht]
\includegraphics[clip,width=10cm,height=7cm]{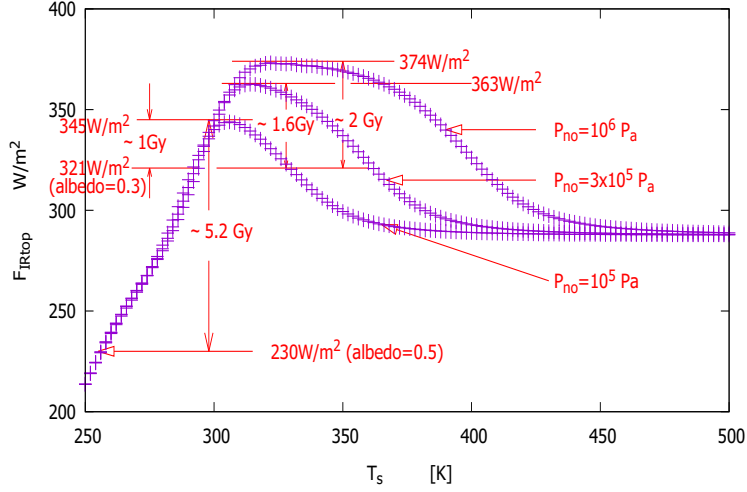}
\caption{Three pressure values of non-condensable component (N$_2$) ($p_{n0}=10^5, 3\times 10^5, $ and $10^6$ Pa) are presented, where 
the runaway critical solar flux is increased for increasing the value of $p_{n0}$. }   
\end{figure}

The results of  3-D simulations are similar to the case of RH $\sim$ 0.45 in the 1-D calculation (Leconte et al. 2013), 
where Relative Humidity(RH) is the ratio of the partial pressure of water vapor to the equilibrium vapor pressure of water at a given temperature. 
Three values of relative humidity (RH=1, 0.6, and 0.45) are presented in Fig. 3, where the runaway critical solar flux is increased for the decreasing value of RH.
Because the optical depth of the atmosphere has decreased for the temperature, the radiation can pass through the atmosphere easily.   
Then  $F_{IRtop}$ is going to increase from the lower atmosphere where the temperature is relatively high.  
Effectively the K-I limit (Komabayashi-Ingersoll limit (Komabayashi 1967; Ingersoll 1969; Nakajima et al. 1992) has increased.  For the cases of RH=0.45 and 0.6, 
the critical solar flux becomes 410W/m$^2$ and 384W/m$^2$, respectively.  
Then  the ocean could exist $\sim$ 3.2Gy and $\sim$ 2.3Gy, because of $(1.00778)^{32.15} \sim 410/321 \simeq 1.2773$ and $(1.00778)^{23.1} \sim 384/321 \simeq 1.196$.
  If we consider cases of albedo = 0.5, the ocean could exist longer than 4.6 Gy.


\begin{figure}[ht]
\includegraphics[clip,width=10cm,height=7cm]{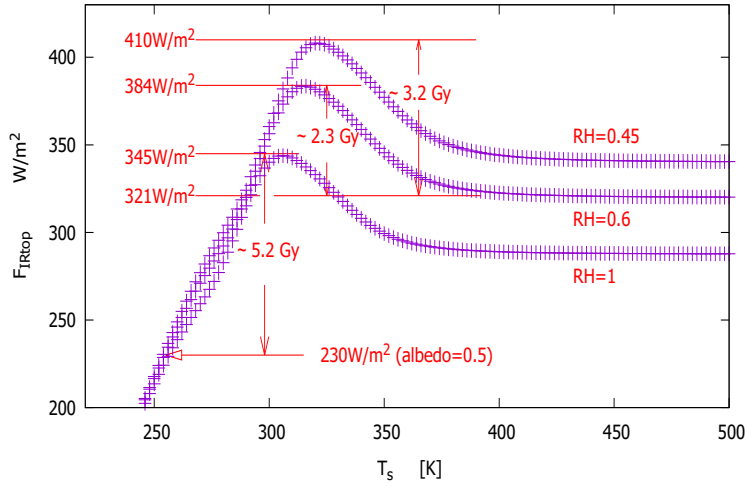}
\caption{Three values of relative humidity (RH=1, 0.6 and 0.45) are presented, where the runaway critical solar luminosity is increased for decreasing value of RH.
Because the optical depth has decreased, the radiation can pass through from the lower atmosphere easily, where the temperature is relatively high. }   
\end{figure}

\begin{figure}[ht]
\includegraphics[clip,width=10cm,height=7cm]{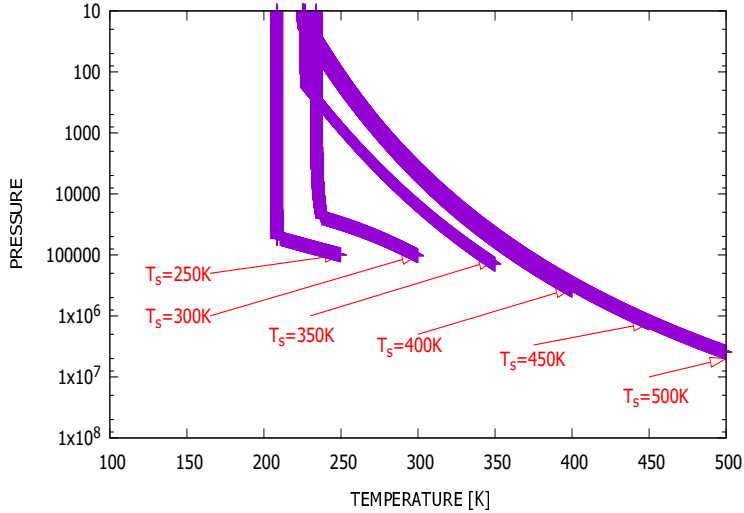}
\caption{Temperature-Pressure relation  for several T$_s$(=250, 300, 350, 400, 450, 500 K) cases are derived.} 
\end{figure}

Temperature and pressure relation for several $T_s$ cases are presented in Fig. 4.  
It could be noticed that temperature becomes the saturated water vapor pressure curve $T=T^*(p)$ when $T_s$ increases.

In Fig. 5, the surface temperature and solar flux relation are presented for Venus and Earth, respectively.  The surface gravitational acceleration  ($g$) is different between Venus and Earth.  
The results could be interpreted that if the atmosphere is compressed by gravity, the temperature of the stratosphere and tropopause could increases.  
Then the escape radiation from the stratosphere and tropopause increases for the large gravity. 
If the assumption could be applied that solar luminosity will increase each $10^8$ y by 0.778$\%$, water ocean on Earth could be expected for $\sim$9.8 Gy 
under this approximation.  
 because of $(1,00778)^{98.4} \sim (358/167) \simeq 2.143$.  For the coming future of Earth, it is important to investigate Venus's history.  
 The abundance of CO$_2$ in Venus is said to be comparable to that in Earth where most CO$_2$ are in rocks through chemical reaction.   
\begin{figure}[ht]
\includegraphics[clip,width=10cm,height=7cm]{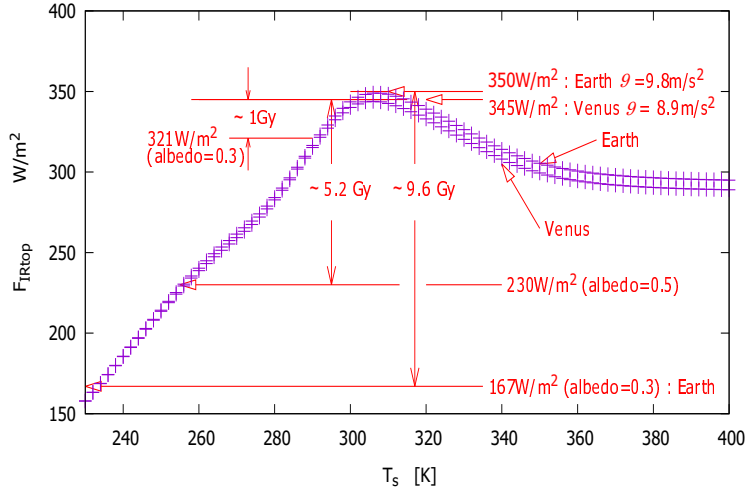}
\caption{Venus vs Earth.  Almost the same for Fig. 1, except the value of $F_{IRtop}(=solar flux)$ on Earth which is a little bit higher than that of Venus.}   
\end{figure}

Mole fraction and pressure relation for several $T_s$ cases are presented in Fig. 6.  
For the cases $T_s > 350$K, mole fraction of condensable component (H$_2$O) becomes dominant.

\begin{figure}[ht]
\includegraphics[clip,width=10cm,height=7cm]{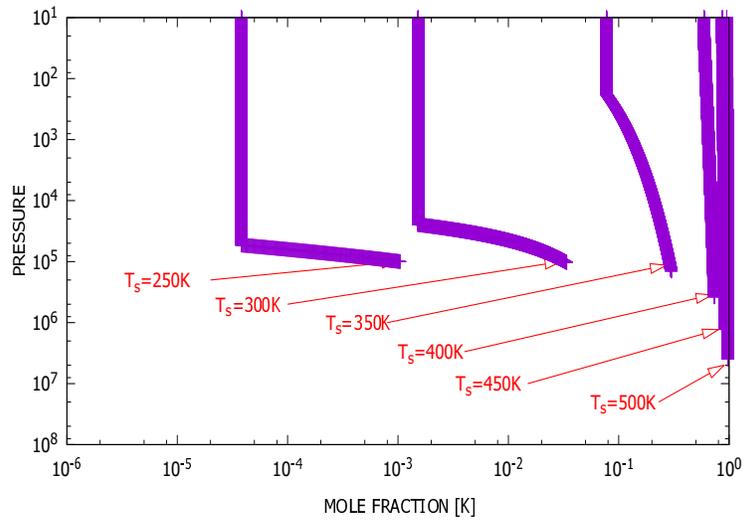}
\caption{Mole Fraction and Temperature relation for several $T_s$ cases for several Ts cases are calculated.}   
\end{figure}

Temperature and optical depth relation for several $T_s$ cases are presented in Fig. 7.  
 For the cases $T_s > 350$K, temperature and optical depth relation becomes the same 
 that of the saturated water vapor pressure curve (Kasting, 1988; Abe and Matsui 1988).

\begin{figure}[ht]
\includegraphics[clip,width=10cm,height=7cm]{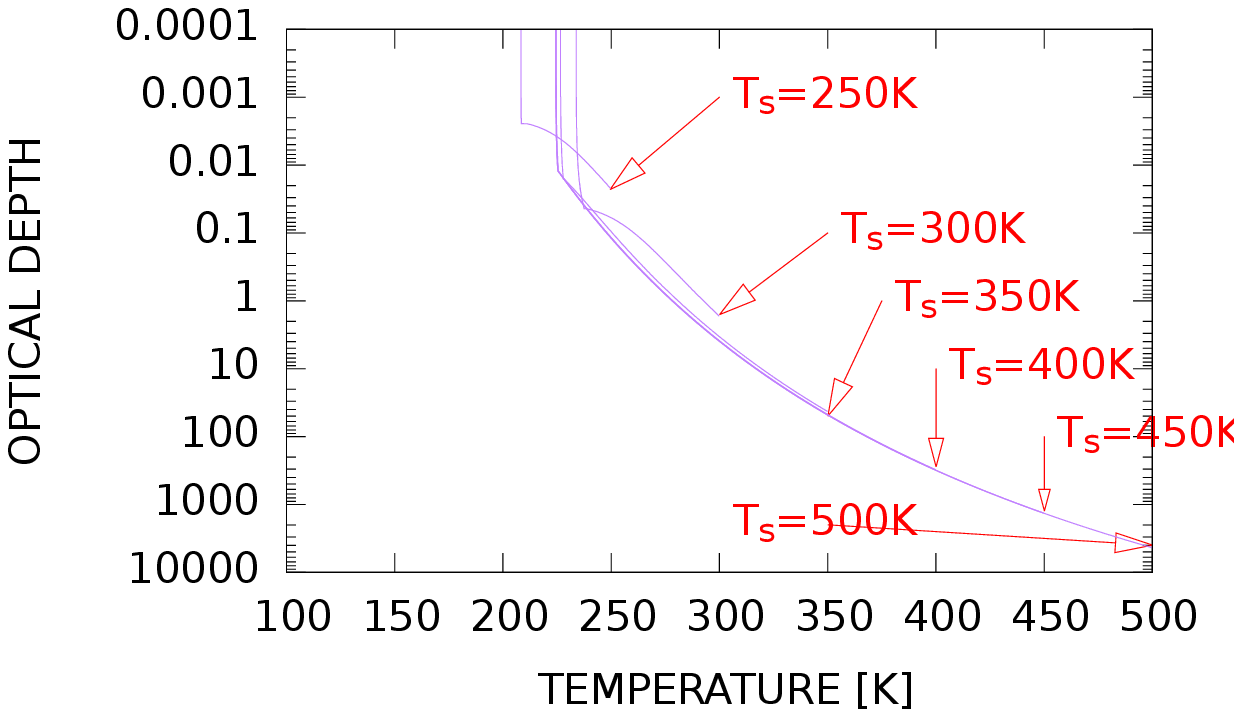}
\caption{ Temperature and Optical Depth relation for several $T_s$ cases are derived.}   
\end{figure}

Mole fraction and optical depth relation for several $T_s$ cases are calculated in Fig. 8.  
$T_{s}$ and $F_{IRtop}$ relations for several initial pressure $p_{n0}$  are presented in Fig. 9.
The Komabayashi-Ingersoll limit is also indicated.  

\begin{figure}[ht]
\includegraphics[clip,width=10cm,height=7cm]{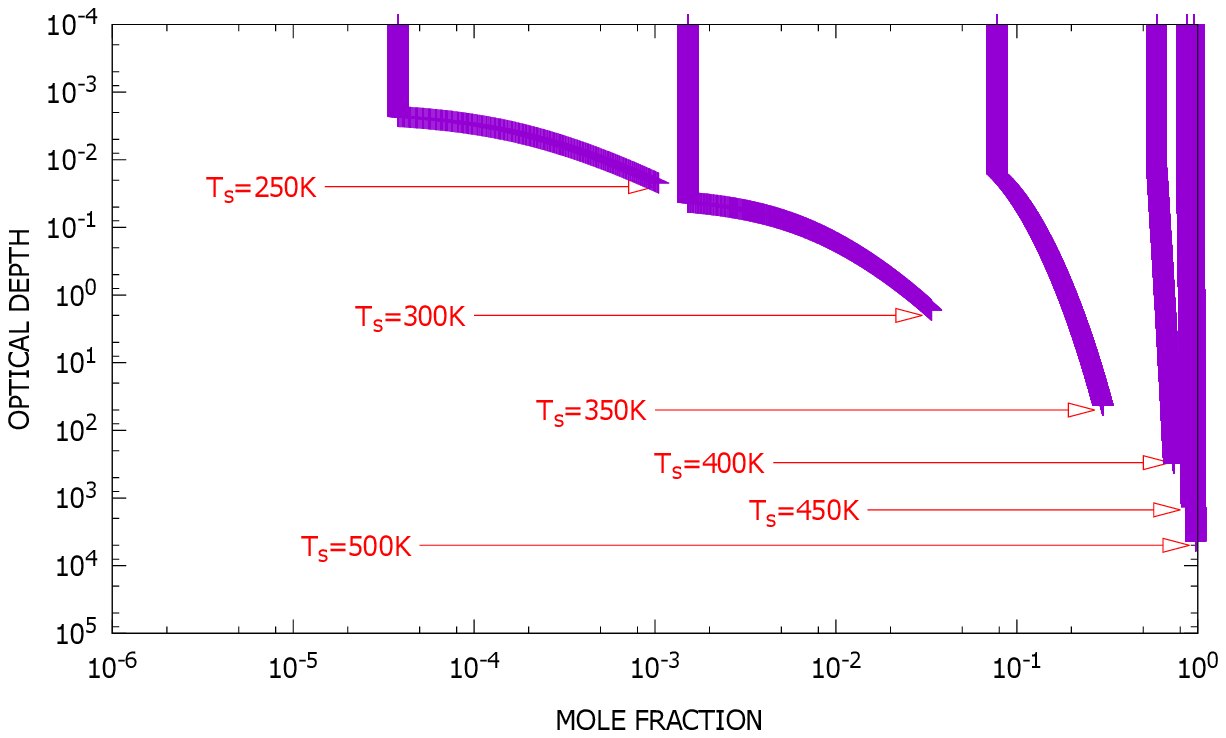}
\caption{Mole Fraction-Optical Depth for several $T_s$ cases}   
\end{figure}

\begin{figure}[ht]
\includegraphics[clip,width=10cm,height=7cm]{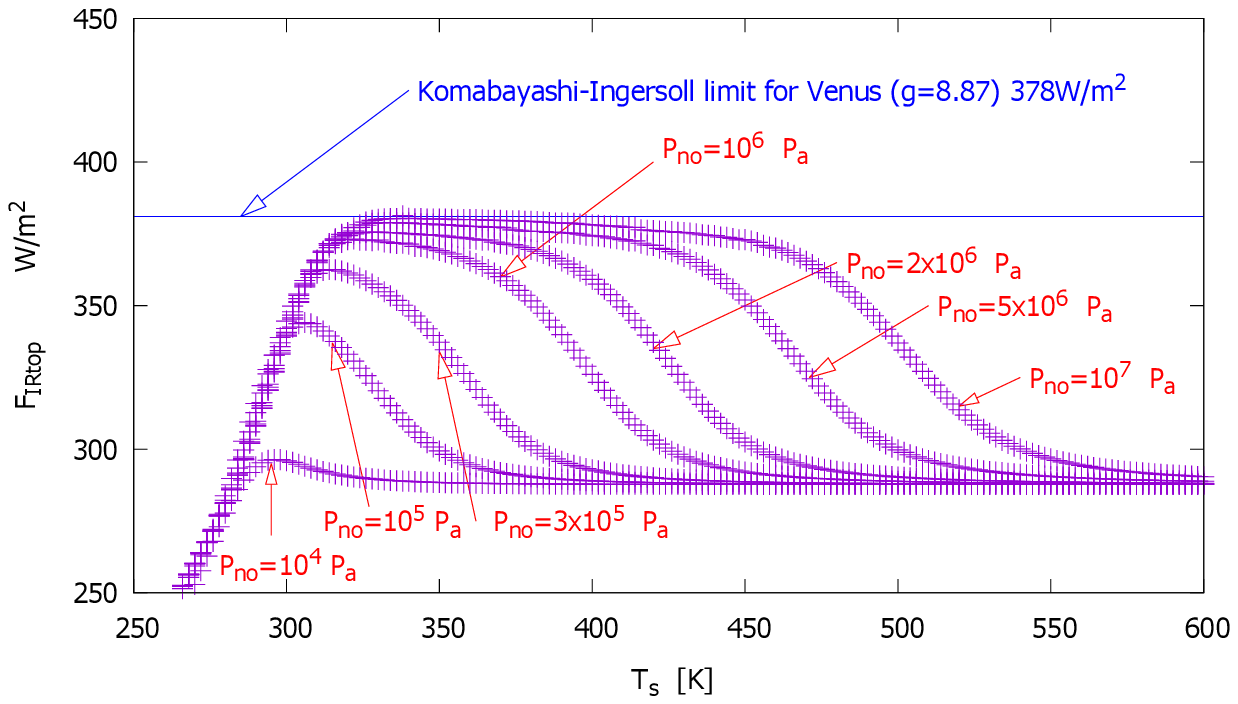}
\caption{The relationships between  $T_{s}$ and $F_{IRtop}$ for several $p_{n0}$ cases are presented}   
\end{figure}

In Figure 10, the radial distance from Sun is shown by log-scale in the horizontal axis with the normalization by Venus distance $r_{\rm venus}$. 
The radiation flux per unit surface is represented by log-scale in the vertical axis.  
Almost 4.6 Gy ago, the primordial solar flux on Venus is 321W/m$^2$ for albedo=0.3 as shown in Figs. 1 and 10.  
If the flux increased to 345W/m$^2$, the runaway would begin. 
Then it is estimated that the ocean on Venus could be expected for almost 1Gy as shown in Fig. 1.

There are two problems pointed by 3-Dimensional calculations (Goldblatt et al. 2013). One is about the albedo (Yang et al. 2014). 
Especially for slowly rotating planets like Venus, the ocean temperature for dayside increases and clouds are formed due to the heated water evaporation.
The reflection by white clouds increases the albedo to almost 0.7 for two times of young faint solar flux (Yang et al. 2014).  
For albedo=0.7, the injection flux would decrease to $321\times(1/0.7)\times0.3 \sim 138$ W/m$^2$.      
This injection flux becomes to 321W/m$^2$ when the distance could decrease to 1/1.53 $\simeq$ 0.65 by Venus distance normalization as (321/138)$^{0.5} \sim $1.53.
Even there it will take almost 1Gy that the injection flux becomes to 345W/m$^2$ when the runaway will begin.
For the critical distance where the injection flux is 345W/m$^2$, the distance could approach to (138/345)$^{0.5} \sim $0.63, 
which is shown in Fig. 10 by the symbol $A$.
The optimistic habitable region could increase to this distance by considering the albedo of slowly rotating planets. 

As Yang et al. (2014) pointed out that the planetary rotation will affect the evolution of the atmosphere.  
Only the distance from the primary star could not determine the habitable region where liquid water could exist.  
It must be considered the albedo including rotational effect.

The other point is noticed by Leconte et al. (2013) that the relative humidity (RH) in the 3D calculation is different from that in the 1D calculation.
The saturated humidity (RH=1) is assumed in the 1D calculation. In 3D calculation characterized by the Hadley circulation, 
the saturated air ascent around the equator and condensate water drops there.  
The air descents adiabatically at a high latitude where the relative humidity decreases. The effective RH becomes 0.45 due to the calculation by Leconte et al. (2013).  
If it is calculated with RH=0.45, the critical runaway flux is increased to 410W/m$^2$ shown in Fig. 3, where the water ocean could be expected for $\sim$ 3.2 Gy.
 
 Yang et al. (2014) also noticed that the relative humidity is low in the slowly rotating planet. 
 They show that the habitable distance from the primary star could be decreased to 0.66 under Venus distance normalization, shown in Fig. 10 around by symbol $A$.

 From the above consideration, the habitable region could increase by 3-D calculation compared to 1-D calculation.



\begin{figure}[ht]
\includegraphics[clip,width=10cm,height=7cm]{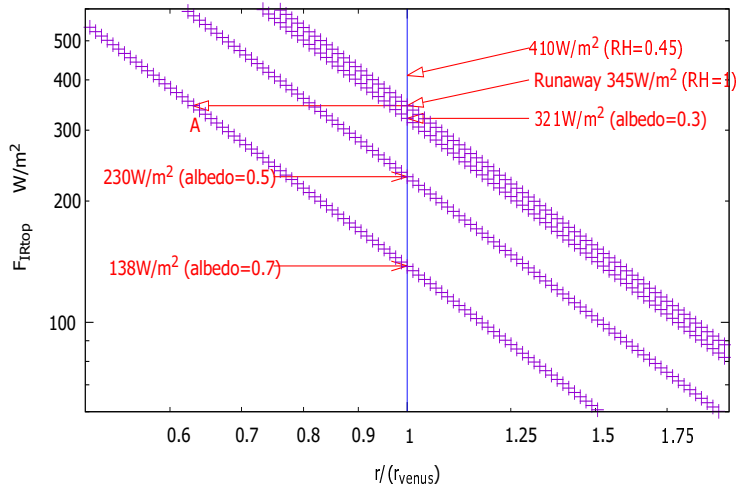}
\caption{The relationships between $F_{IRtop}$ and  distance $r$ normalized by the Venus distance $r_{venus}$. 
Several typical values at $r=1$ are presented.}    
\end{figure}

 \section{Results and Discussion}
 Venus is similar to Earth in size and position in our solar system (Taylor and Grinspoon, 2009). 
It is not certain whether Venus ever had an ocean. 
Venus has a D/H ratio that is about 150 times that of Earth (Donahue et al. 1997). 
It is indicated from the observations by the Venus Express that the planet's highland plateaus were ancient continents surrounded 
by water and made of substances similar to the granite of which formation, it is said, the ocean is necessary (Hashimoto et al. 2008).

An extended habitable period is possible under the traditional estimate, including albedo parameter and rotation period 
(3-Dimensional simulation) (Yang et al. 2014; Way et al. 2016). 
Even 1-Dimensional radiative-convective model (Nakajima et al. 1992) predicts the existence of ocean for $ \sim $ 1 Gy, 
considering the young faint solar flux with modest parameters of albedo=0.3, RH=1, and $p_{n0}$=10$^5$ Pa.  
The 3-Dimensional calculations show the decrease RH and that
slow rotation increases the albedo (Yang et al. 2014; Way et al. 2016).
Venus's climate could have been habitable until at least 0.715 Gya ago 
(chosen to be the most recent resurfacing event on Venus (Way et al. 2016; 2020; Kreslavsky et al. 2015)). 
We have estimated the possible timescale for the ocean history 
under the one-dimensional radiative-convective model, including 3-D calculation results.


a. Various uncertainty

  There are various uncertainty factors to estimate the relationship between the surface temperature $T_{s}$ and 
outgoing infrared radiation at the top of the atmosphere $ F_{IRtop}$.  
Our results do not apply directly to any real planet history because of large uncertainties in our calculation due to the absence of clouds and the use of a one-dimensional model. 
 In order to determine quantitatively, it seems to be necessary to evaluate the parameters such as albedo, effects of clouds, 
 and relative humidity, circulation of the atmosphere over the surface of the planet (Zsom et al. 2013). 
 It will be better to estimate the distribution of those parameters for the greenhouse effects and under the increase of solar luminosity. 

b. Gray opacity approximation

If one considers more than gray opacity approximation, one has to treat opacity, including the line by line treatments, Lorentz factor, pressure effect, Voigt continuous bands, k approximations, random approximations, and further (Pierrehumbert, 2010, Seager, 2010), which make the physical understanding to be complicated.

c. Lump

The cause that the lump appears is explained clearly in the paper (Goldbratt and Watson 2012).  Effective emission to space takes place around $\tau =1$.  
\begin{figure}[ht]
\includegraphics[clip,width=10cm,height=7cm]{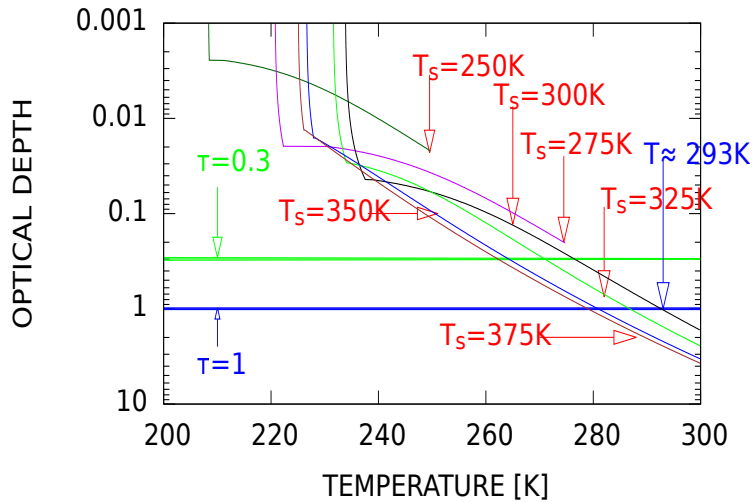}
\caption{Temperature and Optical Depth relation for several $T_s$ cases of $p_{n0}=10^5 $ are presented.}   
\end{figure}

The detailed example is shown in Fig. 11.  Temperature and Optical Depth relation for several $T_s$ cases of $p_{n0}=10^5 $ are presented there.  This figure is almost the same as Fig.7, however, the curves are multiplied and enlarged.  Especially one must pay attention to the temperature that each curve is crossing the line (blue) of constant $\tau $=1.  It is related to the lump height of Fig. 1.  For example, the curve of T$_s$=300K (black) crosses the $\tau =1$ line at around T$\simeq$ 293K which is the highest temperature and corresponds to the top of the lump of Fig. 1.  This is just what means "Effective emission to space takes place around $\tau =1$. "  Another example is the curve of T$_s=275K$ (purple) that does not reach the $\tau =1$ line, meaning that the outgoing flux is not enough from the atmosphere.  Some part of the flux must be emitted from the surface.   
      
The other example is the curve of T$_s=325K$ (green) which crosses the $\tau =1$ line at around T$\simeq$ 283.  It is lower than T$\simeq$ 293K , then the outgoing flux is smaller than the top of the lump as in Fig. 1.

The presence of background gas decreases the water vapor mixing ratio, increasing the moist adiabatic lapse rate.  The temperature around an optical depth of unity is higher and more radiation can be emitted than the cases of lower background pressure and pure water vapor atmosphere.    

\begin{figure}[ht]
\includegraphics[clip,width=10cm,height=7cm]{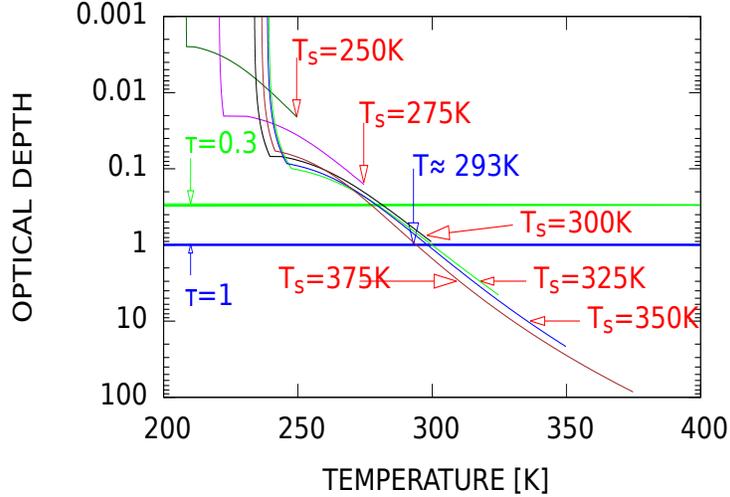}
\caption{Temperature and Optical Depth relation for several $T_s$ cases of $p_{n0}=10^6$  are derived.  This figure is almost analogous to Fig.7 and Fig. 11, however the background pressure is $p_{n0}=10^6 $.  One must note that the temperature of each curve, crossing the line (blue) of constant $\tau $=1, comparing the lump height of the curve of $p_{n0}=10^6$ on Fig. 9.  The curve of T$_s$=325K (green) crosses the $\tau =1$ line at around T$\simeq$ 300K which is the highest temperature on Fig. 12 and corresponds to the almost top of the lump of the curve of $p_{n0}=10^6$ on Fig. 9. 
Another curve of T$_s=350K$ (blue) crosses the $\tau =1$ line at around T$\simeq$ 300  which is almost the same as the case of T$_s=325K$ (green) and the outgoing flux is almost the same as $p_{n0}=10^6$ on Fig. 9. 
Another curve of T$_s=375K$ (brown) crosses the $\tau =1$ line at around T$\simeq$ 293  which is lower than T$\simeq$ 300K and the outgoig flux is smaller than the top of the lump as in the case of $p_{n0}=10^6$ on Fig. 9.}
\end{figure}

The example is shown in Fig. 12 for the case $p_{n0}=10^6 $ and the explanation of the figure is presented in the figure caption.

However, there is no lump in the paper by Kasting et al. (1988) which has paid much attention to the variation of albedo, CO$_2$ pressure, and H$_2$O absorption coefficient.  The results of Fig. 1 (a) in the paper by Kopparapu et al. (2014), including the albedo effects, show some lump for the case of $p_{N2}$=10bar.  However, the results of varying N$_2$ pressure seem to be utterly different from the results of Fig. 6 by Nakajima et al. which do not consider the variation of the albedo.

  The gray opacity approximation seems to be not the cause of the lump disappearance.  In the text of Pierrehuumbert (2011, Fig. 4.38), there are lumps for the cases for 100 mb N$_2$, 1 bar N$_2$, and 2 bar N$_2$ for g=20m/s$^2$, noting that calculation was done with the homebrew exponential sums radiation code, incorporating both 1000cm$^-1$ and 2200cm$^-1$ continua.  

There seems to be no clear explanation for the lump disappearance if one considers non-gray simulation.

\subsection{Possible Venus History-Conjecture}

At present Venus is hot ($\simeq $730K), dry, and has no ocean.
No plate-tectonics (tectonic feature) is remained (Tessera).  It is inferred that
 the lava volcanic re-surfacing around last $\sim $1Gy ($\sim$ 0.715 Gy) ago.  
 It could be imagined the following as the evolutionary history of the water ocean on Venus 
for optimistic conjecture (the following $X, Y$ and $Z$ might be numbers, indicating uncertainties).
At the initial time, magma ocean may exist and later heavy bombardment event occurred $\sim X \times 10^8$ y.
After then Venus becomes cool down and heavy rain makes the ocean in the low land.
 There is a possibility that the mild climate could exist $\sim Y \times 10^9$ y. 
 During this time plate-tectonics may work.  
 Lava igneous provinces (LIPs) would be formed during this era (Kreslavsky et al. 2015).

Last $(Z \sim 4.6) $Gy, multiple large-scale eruptions and lava volcanic resurfacing would have occurred.
The large amount of CO$_2$ would have degassed and its greenhouse effect would have warmed the atmosphere.  
Greenhouse Runaway event might occur almost $10^9$ y ago.
Then the ocean evaporates.

Under a modest parameters of albedo=0.3, RH=1 and $p_{n0}$=10$^5$ Pa,
 there is a possibility of the water ocean on Venus for $\sim$ 1Gy.  
 If we consider 3-D calculations showing albedo $>$ 0.5 and RH $<$ 1, 
 it could be expected that the ocean could exist longer than $\sim$ 4.6 Gy under the solar luminosity increasing.
  If then, there must be another cause such as huge volcanic emission with global lava resurfacing.

It is important to investigate Venus's history for the coming future of Earth 
and observations of exoplanets for their habitable zones.
\vspace{3mm}


{\normalsize Table I}\hspace{5mm}{\normalsize  Conjecture: Evolutionary history of the water ocean on Venus}
\vspace{-1mm}
\begin{center}
\begin{tabular}{|c|l|} \hline 
 $\sim X \times 10^8$  y  &    Magma \  Ocean, Heavy \  Bombardment.\\ \hline
 X  $\times 10^8  \sim 10^9$   y   &    Heavy \  Rain, Water \  Ocean \  Formed,  CO$_2$ \   \  dissolution. \\ \hline
$ 10^9 \sim Y \times 10^9 $  y  &    Mild \  climate, Ocean \ and  \  Plate \  Tectonics \ exist.\\  
          &  Tessera \  (highly \ deformed \ terrain ) \ are \ formed. \\ \hline
$  Y \times 10^9 \sim  Z \times 10^9 $  y  &  Lava \ volcanic \ resurfacing \ (Ocean \ evaporate). \   Released  CO$_2$ .\\                                         
                   &   Runaway \  Greenhouse \ Effect \ (Temperature \ increased). \\ 
                   &  Plate \ Tectonics \ end.  \\ \hline
 $  Z  \times 10^9 \sim 4.6  \times 10^9 $  y   &     CO$_2$ \   clouds \ are \ formed.  \  Present.\\  \hline
\end{tabular} 

\vspace{5mm}

\end{center}



Acknowledgments

\vspace{2mm}
We  would like to thank  M. Takagi and  H. Sagawa for stimulating suggestion.  
A. S. would like to thank T. Sasaki for valuable discussions.


\vspace{0.4cm}

\end{document}